\documentclass[submission,copyright,creativecommons]{eptcs}
\providecommand{\event}{VPT 2017}

\usepackage{breakurl}             % Not needed if you use pdflatex only.
\usepackage[utf8]{inputenc}
\usepackage{color}
\usepackage{caption}
\usepackage{enumitem}
\usepackage{listings}
\usepackage{amsmath}
\usepackage{amsfonts}
\usepackage{multicol}
\usepackage[all]{xy}
\usepackage{mathtools}
\usepackage{amssymb}

\newcommand{\cell}[2]{\langle #2 \rangle_\mathsf{#1}}
\newcommand{\cellpre}[2]{\langle #2 \ellipses \rangle_\mathsf{#1}}

\newcommand{\bigcell}[2]{\left\langle #2 \right\rangle_\mathsf{#1}}

\newcommand{\term}[1]{\, \texttt{#1}\,}
\newcommand{\var}[1]{\, \textit{#1}\,}
\newcommand{\varType}[2]{\, \textit{#1} : \textit{#2}\,}

\newcommand{\ellipses}{\mathrel{\cdot\!\!\cdot\!\!\cdot}} 
\newcommand{\kdot}{\bdot}
\newcommand{\bdot}{{\scriptscriptstyle\bullet}}
\newcommand{\dotCt}[1]{%
\ensuremath{{\raise.3ex\hbox{\ensuremath{\kdot}}}_{\color{black!60}\scriptstyle\it \!#1}}}
\newcommand{\khole}{\Box}
\newcommand{\kra}{\curvearrowright}

\newcommand{\sms}[2][]{%
\small%
\begin{equation}\label{#1}%
#2
\end{equation}%
\normalsize}%

\newcommand{\smsnn}[1]{%
\small%
\begin{equation*}%
#1
\end{equation*}%
\normalsize}%

\title{Trustworthy Refactoring via Decomposition and Schemes:\\ A Complex Case Study}

\newcommand{\elte}{Eötvös Loránd University}

\author{
Dániel Horpácsi \institute{\elte} \email{daniel-h@elte.hu}
\and
Judit Kőszegi \institute{\elte} \email{koszegijudit@elte.hu}
\and
Zoltán Horváth \institute{\elte} \email{hz@elte.hu}
}
\def\titlerunning{Trustworthy Refactoring via Decomposition and Schemes}
\def\authorrunning{D. Horpácsi, J. Kőszegi and Z. Horváth}

\usepackage{tikz}

\makeatletter
\newenvironment{btHighlight}[1][]
{\begingroup\tikzset{bt@Highlight@par/.style={#1}}\begin{lrbox}{\@tempboxa}}
{\end{lrbox}\bt@HL@box[bt@Highlight@par]{\@tempboxa}\endgroup}

\newcommand\btHL[1][]{%
  \begin{btHighlight}[#1]\bgroup\aftergroup\bt@HL@endenv%
}
\def\bt@HL@endenv{%
  \end{btHighlight}%   
  \egroup
}
\newcommand{\bt@HL@box}[2][]{%
  \tikz[#1]{%
    \pgfpathrectangle{\pgfpoint{1pt}{0pt}}{\pgfpoint{\wd #2}{\ht #2}}%
    \pgfusepath{use as bounding box}%
    \node[anchor=base west, fill=orange!30,outer sep=0pt,inner xsep=1pt, inner ysep=0pt, rounded corners=3pt, minimum height=\ht\strutbox+1pt,#1]{\raisebox{1pt}{\strut}\strut\usebox{#2}};
  }%
}
\makeatother

\begin{document}
\maketitle

\begin{abstract}
Widely used complex code refactoring tools lack a solid reasoning about the correctness of the transformations they implement, whilst interest in proven correct refactoring is ever increasing as only formal verification can provide true confidence in applying tool-automated refactoring to industrial-scale code. By using our strategic rewriting based refactoring specification language, we present the decomposition of a complex transformation into smaller steps that can be expressed as instances of refactoring schemes, then we demonstrate the semi-automatic formal verification of the components based on a theoretical understanding of the semantics of the programming language. The extensible and verifiable refactoring definitions can be executed in our interpreter built on top of a static analyser framework.
\end{abstract}

% ------------------------------------------------------------------------------

\section{Introduction}

Refactoring is a widely used technique that improves non-functional properties of source code without affecting dynamic semantics. Tool-assisted refactoring transformations must be trustworthy if programmers are to be confident in applying them on arbitrarily extensive and complex code in order to improve style or efficiency. Unfortunately, currently available, widely used refactoring frameworks do not guarantee the correctness of the transformations they implement, thus application of these might introduce undesired bugs to the system. In an ideal setting, such frameworks should provide formal guarantees, and they also should be extensible, so that programmers can compose their own program transformations according to their need. In this paper, we demonstrate our latest results in defining trustworthy refactoring definitions, with particular attention to a fairly complex case study which is intended to demonstrate the applicability of our approach to verifiable refactoring definition. We defined the refactoring formalism for the functional programming language Erlang, but we did so with other languages and paradigms in mind; the presentation is as independent of the functional object language as possible.

The main contributions of this paper are:
\begin{itemize}
\item Definition and application of language-agnostic refactoring schemes,
\item Decomposition of a complex refactoring transformation into a series of prime refactorings,
\item Definition of the prime refactorings in terms of instantiations of refactoring schemes,
\item Demonstration of the verifiability of refactoring schemes and scheme instances based on the formal semantics of the object language.
\end{itemize}

\noindent
The rest of the paper is structured as follows. Section~\ref{sec:back} summarises the background of the work and at the same time it gives a brief overview of related results. Section~\ref{sec:app} explains the core ideas behind our approach to refactoring definition and verification, Section~\ref{sec:dec} presents the definition of a complex case study in our formalism, and then Section~\ref{sec:verification} demonstrates the verifiability of the case study. Finally, Section~\ref{sec:conc} concludes.

\section{Background and related work}
\label{sec:back}

In classic terminology, refactoring is a behaviour-preserving program transformation. From a more theoretical viewpoint, refactoring is a transformation on some model of the program, where applying the transformation to the model of a particular program should result in a semantically equivalent model with respect to some definition of formal semantics and semantic equivalence.

\subsection{Program representation}

There are several ways to represent the syntax and static semantics of program source code, but in most cases, an extension of the abstract syntax tree (AST) is used. This is beneficial for at least two reasons: 1) the syntax is easiest captured and manipulated by using the syntax tree, and 2) the formal semantics of programs is usually defined on the AST, thus any models that include (or can be turned into) the AST are amenable to semantic equivalence checking. Formal semantics usually captures the operational behaviour of the code, while semantic equivalence states that for any input, the original and resulting programs produce the same output.

The role of semantic program models is to capture the syntax and static semantics of the code in a compact yet efficient way. This is usually achieved via static semantic analysis~\cite{cefp11lecture} that enriches the abstract syntax tree with semantic nodes and edges representing context-sensitive properties. A good program representation simplifies code understanding, further static analysis as well as program transformation. The definition of a refactoring transformation highly depends on the abstraction level and richness of the used program model: the more details the model captures, the easier to express preconditions and transformation steps. In the ideal case, the model uses the same abstractions as the refactoring specification.

\subsection{Transformation definition}

Essentially, the interpretation of a refactoring transformation is a deterministic relation from programs to programs, or more precisely, from program models to program models. When the program model is the abstract syntax tree, this relation can practically be described with (strategic) term rewriting. In case the model happens to be a graph, graph rewriting can be applied instead. Indeed, the two classic approaches to formally defining refactoring are term rewriting and graph rewriting, and different refactoring specification formalisms adapt the core ideas of rewriting.

With the emerging use of tool-assisted refactoring, the interest in specification and verification of transformations has been constantly increasing and guarantees are needed for users to trust the complex changes carried out by tools. The abstractions for defining refactoring as well as the levels of correctness guarantees are varying, but almost every approach incorporates the fundamental work of Opdyke~\cite{Opdyke:1992:ROF:169783} that suggests refactorings be composed of basic steps called micro-refactorings. Simpler transformations are easier to read, write and to verify, but on the other hand, decomposition of extensive refactorings to simple steps requires experience and considerable effort.

\paragraph{Strategic term rewriting.}

Context-free conditional rewrite rules and functional strategies~\cite{Bravenboer200852} are widely used to implement program transformations and stuctured data transformations in general. Furthermore, Bravenboer and Olmos show that by adding dynamically defined rewrite rules into the system~\cite{Bravenboer:2005:PTS:1227247.1227253}, context-dependent transformations~\cite{Olmos05} are also definable. 

Beyond any questions traversal programming is an expressive and exciting paradigm, but as Lämmel and others~\cite{strategyerrors} point out in their comprehensive study, error-free use of strategy combinators requires expertise, not mentioning the difficulties of formal verification (the termination property of a complex strategy alone is a considerably difficult problem). The paper characterises typical mistakes in strategic programming, and one of their findings is that errors mostly stem from mixing up selection of terms of interest (their type and pattern), keeping track of the origin of data, checking side conditions and doing actual transformation.

\paragraph{Graph rewriting.}

Semantic program graphs capture the binding structure, the data and control flow relations in the program, while they may also depict properties of specific program symbols. It is apparent that semantics-aware, verifiable transformations can be specified with graph rewriting~\cite{SMR:SMR316} as well, but the graphical descriptions of graph rewrite rules are relatively complex compared to concrete syntax patterns. In addition, matching a graph pattern to a semantic program graph is computationally more complex than matching a first-order term pattern to a term. Since the graphical format of rules is representation-dependent and rather complicated, this system is less likely to be used by users to define their own refactorings. Some systems use a graph model, but express the context-sensitive rewrite rules with a special textual representation, e.g. Padioleau et al.~\cite{Padioleau:plos06} use a transformation language incorporating semantic conditions into the textual patterns.

\paragraph{Refactoring languages.}

Designing domain specific languages for refactoring programming is a well-known idea, there are related results for different object languages with different representations. Some of these define the entire code transformation logic including term-level rewriting, while some only offer a formalism for composing atomic steps in a convenient way. Leitdo~\cite{Leitdo} gives an executable, rewrite-based refactoring language with expressive patterns,  Verbaere~\cite{Verbaere:2006:JSL:1134285.1134311} proposes a compact, representation-level formalism for executable definitions. These formalisms are expressive and language-independent, but at this generality they cannot support correctness checks for refactoring definitions. For Erlang, Li and Thompson~\cite{Li:2012:DLS:2259278.2259323} define an API for describing prime refactorings and a feature-rich language for interaction-aware composition, but formal verification is not addressed in their work either.

\subsection{Verified refactoring}

For object-oriented languages, Schaefer~\cite{Schaefer:2010:SIR:1932682.1869485} introduced a system in which he reasoned about semi-formal definitions of a set of basic refactorings. The idea of using locks and language extensions instead of preconditions is exciting and promising, but the expressiveness is limited due to the lack of custom rewritings, and the proofs are mostly informal rather than formal. Roberts~\cite{Roberts:1999:PAR:929806} applies a different definition style, with an emphasis on the side-conditions and proper composition of the base refactorings. However, neither of them provides formally verified or executable definitions. There are some results~\cite{ver-ref-sultana} in defining provably correct refactorings for simple languages, some mechanised proofs even for real-world refactorings~\cite{julien} are available, but none of these allow for defining custom transformations and provide automatic verification for those.

We cannot talk about verified refactoring without clarifying object language semantics and the definition of semantic equivalence. Matching logic and reachability logic provide a formal system~\cite{sem-based-verif} for defining formal semantics for programming languages, doing model checking, and also for expressing semantic equivalence. Although they demonstrate how to use the framework for proving equivalence of programs (and program patterns), they do not employ the system for checking transformation definition correctness.

\section{Our approach: decomposition to refactoring schemes}
\label{sec:app}

In our prior work~\cite{VPT16}, we introduced novel abstractions for executable and verifiable refactoring definitions. Besides incorporating the basic idea of decomposition to micro-refactoring, we use a semantic program graph representation to define transformations in terms of strategic term rewriting rules extended with language-level semantic conditions. Unlike strategic programs, our definitions separate analysis, target selection (object or term of interest), side condition checks and actual transformation. Our transformation language includes the well-known basic strategies, but in order for a transformation to be verifiable, it has to be expressed with the semantic schemes we provide for refactoring programming. These schemes depict special patterns in the program graph (selection), define some preconditions for the refactoring (condition), and control the rewrite rules that instantiate the scheme (transformation). Semantic schemes can be seen as verified algorithmic skeletons (or rewrite strategies), whose instances can be automatically verified for correctness.

\subsection{Prime and composite refactoring functions}

Refactorings are defined with functions, either prime (specifying a non-decomposable refactoring step) or composite (composing refactoring steps together). Refactoring functions have a target (the object of interest) determining a reference to a syntactic or semantic element in the program, and a return value characterising the result. Refactoring functions can take an arbitrary number of arguments of primitive types or node references, and they can define local variables by matching or with binding conditions.

Each prime refactoring function is an instance of a refactoring scheme, which provides a declarative description of the desired change: control is encoded into the interpretation of the scheme, while local variables in prime functions are all single-assignment. On the other hand, composite functions are imperative descriptions controlling application of other refactoring functions with target selectors (query functions for finding program elements of interest), sequential composition and iteration.

\subsection{Schemes}

Extensive code transformations can be expressed with traversal strategies, strategy combinators and complex semantic queries. Nevertheless, our goal is to make every transformation not only executable but also automatically verifiable. To achieve this, we hide semantics-based conditions and control behind pre-verified schemes. They can be understood as complex strategies in traversal programming, but in fact they are much more: schemes define the format of their parameter rewrite rules and may inspect the elements of the rules in order to define the compound strategy they carry out.

Since complex data and control dependencies are present among the various elements of the program, some transformations can only be correct (i.e. behaviour-preserving) if all the dependencies are handled properly when the origin of the dependency changes. Schemes make sure that the transformation will reach out to all code locations that might be affected, and also make sure that the changes made are consistent. Intentionally, schemes hide the complexity connected to semantics-based term selection and side conditions, while at the same time they fully control the application of rewrite rules by relying on these semantic connections. Schemes are instantiated with a series of conditional term rewrite rules, which are expressed in the concrete syntax of the object language, and they may refer to pre-defined semantic functions and predicates. These latter provide access to the program representation with an interface that resembles object language level concepts, allowing anyone knowing the object language to read and write refactoring definitions.

\paragraph{Local refactoring.}

The simplest scheme transforms a single sub-tree (or sub-term) in the program, and there is no control or conditions built into this strategy. Local refactoring simply applies the rewrite rule it takes directly to the program element selected for transformation.

\vspace{-0.5em}
\paragraph{Data-flow and control-flow driven refactoring.}

One of the core ideas of schemes is that dependencies connect program elements that shall be changed consistently. Data-flow induces data dependency, so when an element of a data-flow chain is changed, it entails the need for adjusting the rest the chain. We have two schemes that can be used for refactoring data-flow chains: \emph{forward data-flow}, which starts from the data origin and visits references, and \emph{backward data-flow}, which first modifies the data reference and then compensates data sources accordingly to keep consistency.

\vspace{-0.5em}
\paragraph{Binding driven refactoring.}

Names can induce data as well as control dependencies, and in most cases, when changing binding definitions, references have to be adjusted in order to preserve behaviour. Since our case study object language is Erlang, we identified refactoring schemes for \emph{refactoring variables, functions, records and types}. Any semantic objects that can be given a name can be treated the same way, and obviously, in different programming languages, the set of these will differ.

\vspace{-0.5em}
\paragraph{Introduce binding.}

Introducing abstractions into the program is special in some sense, because although it involves changes at two different locations, one is merely addition and only the other is modification. Schemes of this kind introduce a name and, at the same time, they rewrite a piece of code to use the new binding. In fact, the change is the use of a name, and the compensation of this change is introducing the binding. Semantically, not only the binding is added to the code, but inherently the flow and dependency graphs are extended, too. Currently, we have schemes that \emph{introduce variables and functions} by extracting expressions.

\subsection{Correctness}

We express refactoring correctness in terms of a set of equivalence formulas. In our previous work~\cite{VPT16} we presented how to use a proof system to prove refactorings whose correctness can be expressed by the equivalence of two expression patterns under a given condition. We reduced the equivalence property of the two expression patterns to a correctness property of an aggregated program constructed by the two expression patterns (according to~\cite{eq-corr}), then we applied the language-independent, general-purpose proof system to automatically check the validity of our property.

According to our terminology, refactoring correctness is defined with respect to a formal semantics of the object language and an equivalence relation. We formalised a nearly complete, sequential and deterministic sub-language of Erlang with matching logic formulas, used throughout our proofs. Although the presentation of the entire language definition is beyond the scope of this paper, we include the semantic rules we apply when demonstrating the case study verification. We also define a suitable equivalence formula for our proofs in Section~\ref{sec:verification}.

In order to verify refactorings, we turn refactoring functions into sets of conditional equivalence formulas. For local refactorings, this means simply treating the conditional rewrite rule as a pair of patterns; for strategy-combined rewritings, we face a more complex issue that has to glue rewriting, context and control. We split the verification problem in half: check that the scheme is correct under some assumptions (i.e. a contract), and then prove that the instantiation of the scheme satisfies those assumptions. Typically, contracts are equivalence formulas constructed from elements of the instantiation rules, while the verification of the scheme itself is a structural induction proof with base cases proven by the contract.

\section{Decomposition and definition of the case study}
\label{sec:dec}

The methodology of applying decomposition and schemes for defining refactoring is best demonstrated through a meaningful case study. We explain the decomposition process and the role of schemes as building blocks by formally specifying a well-known and fairly complex function refactoring: \emph{generalise function definition}. As object language, we use Erlang~\cite{CesariniThompson}, an impure, eagerly evaluated, dynamically typed functional programming language.

\subsection{Informal specification}

Our case study ``generalise function'' is a refactoring transformation that turns some value (i.e. a sub-expression within the function body) into a function parameter, thus making the function more abstract. The generalization increases the function arity by one, meaning it will take an extra formal argument compared to the original signature --- this requires that this generalised signature is not defined in the code yet, which is one of the side-conditions of this transformation. In practice, there are two well-known realisations of this refactoring:
\begin{enumerate}
\item Generalise the function and then create a fall-back version with the original arity, where the fall-back function simply invokes the newly generalised version by passing as extra argument the extracted expression. This way, call sites to the original function can be left unchanged, since by calling the fall-back function their behaviour remains the same.
\item Change the call sites so that they pass the extracted expression as extra argument to the new, generalised function. This variant does not duplicate the function, but may affect a large number of code locations if there are a number of references to the generalised function.
\end{enumerate}

\noindent
The first variant is more local as the effect of the transformation remains in the module, while the second variant might reach out to other modules calling the generalised function. In both versions, the expression in question is moved from the function body to the call sites, thus the transformation has to make sure that the binding structure present in the expression is not affected by the relocation. Also common in both variants that they refactor variable and function objects in a general manner, which makes their definition pretty similar. In fact, the first one is a bit more challenging as it both changes the original function and adds a new one, which have to be kept semantically consistent, so we put our focus on defining the first variant of the refactoring.

\paragraph{Example.}

In order to demonstrate the behaviour of this transformation, we present a small piece of code and generalise the function $f$ by lifting the constant $2$ into a function argument. The presented example is intentionally overly simple, yet it shows how the abstractions are extended and changed, which sheds some light on what kind of schemes might be needed for ensuring consistent modification.

\begin{lstlisting}[caption=Original code]
f(X) -> begin X * 2 end.         % function to be generalised
g(X) -> f(X+1).                  % a reference
\end{lstlisting}

\noindent
The refactoring generalises the function by adding a new parameter to it and replacing the constant value with the new parameter in the body. At the same time, it creates a copy of the function that simply calls the generalised one with the original constant value. It might seem useless in this example, but because the expression we relocate may have side-effects, it should get encapsulated by a lambda function (denoted with the \texttt{fun} keyword in Erlang) and its application --- this encapsulation enables the refactoring to keep the order and number of side-effects.

\begin{lstlisting}[caption=Refactored code]
f(X, Y) -> begin X * Y() end.    % new, generalised function
f(X)    -> f(X, fun() -> 2 end). % invokes the new one
g(X)    -> f(X+1).               % callee unchanged
\end{lstlisting}

\noindent
After carrying out function generalization, new names and signatures appear: a ``new'' $f$ taking two arguments gets introduced, where the last argument is the newly introduced variable that takes the extra function parameter. In the next section, we elaborate on how the introduction and manipulation of these abstractions can be split into multiple stages.

\subsection{Decomposition}

By decomposition, we mean expressing a complex refactoring transformation in terms of smaller, simpler refactoring steps. This requires additional effort compared to specifying a transformation as a whole, but it pays off: smaller steps are easier to read, write, and are more easily checked for correctness. In order to decompose a refactoring, we need to understand how it affects language objects, data-flow and control-flow, clarifying how it is boiled down to simpler yet behaviour-preserving steps. Note that in many cases, there are multiple possible decompositions, which may differ in complexity and verifiability.

\paragraph{Avoiding detached refactoring.}

When designing decomposition, we avoid hidden or detached changes, i.e. those that introduce or modify dead code. These are easy to reason about since they are not part of the control-flow nor the data-flow (their modification is not observable from the semantic point of view), but relating detached changes to the original program requires overly complex syntactic or semantic conditions. In the most difficult case, side conditions involving dynamic semantic equivalence of arbitrary expressions might be needed, which we do not support in our formalism. As a matter of fact, we do not incorporate the formal semantics of the object language into the refactoring execution. When checking equivalence is inevitable, the condition might refer to a more restrictive condition that ensures syntactic equivalence.

The ``generalise function'' refactoring could be seen as two big, standalone steps: a (detached) refactoring that creates the generalised function definition, plus another one, which rewrites the original function as an application of the generalised one. Needless to say, this would pose a need for a complex precondition for the second step, namely a formula ensuring that calling the generalised function with the originally selected expression as extra argument is semantically equivalent to the original function body. Rather than composing the complex transformation of two independent transformations, we are going to specify it as a composition of several refactoring scheme instances.

\paragraph{Scenario.}

By building on the refactoring schemes identified in the previous section, we divide the case study refactoring into prime refactoring transformations that are easier to understand and verify. It is apparent that the complex refactoring will introduce new abstractions: a new function abstraction is created for the generalised instance, and a variable abstraction is created for the new parameter holding the value of the generalised expression. Rather than copying the function and then inlining, or adding an unused parameter and then integrating it into the body, we operate with slight yet completely behaviour-preserving changes to the abstractions. In each step, we highlight the term of interest we rewrite with a micro-refactoring (also, on the arrows, we identify the refactoring function and its arguments).

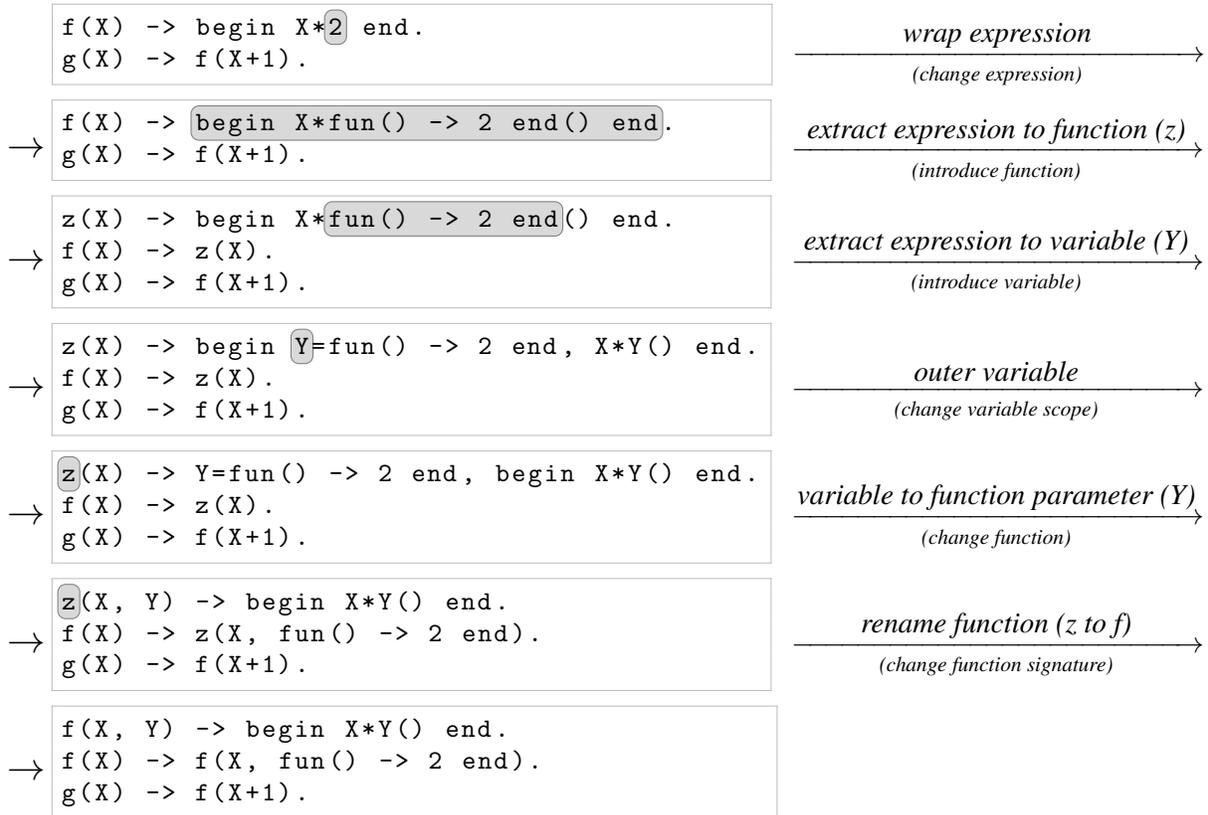
\vspace{0.5em}

\lstset{
  frame=single,
%  framerule=1px,
  rulecolor=\color{lightgray},
  language=Erlang,
  basicstyle=\small\ttfamily,
  breakindent=4ex,
  postbreak=(cont.),
  tabsize=2,
  breaklines=true,
  showspaces=false,
  showtabs=false,
  xleftmargin=30pt,
  framexleftmargin=3pt,
  captionpos=b,
  morekeywords={fun,begin,lists,map,filter,foreach,foldl,skel,do},
  moredelim=**[is][{\btHL[fill=gray!30,draw=gray]}]{@}{@},
}

\noindent
\hspace{17.5pt}
\begin{minipage}{.584\textwidth}
\begin{lstlisting}[frame=single,xleftmargin=0pt,framexleftmargin=0pt]
f(X) -> begin X*@2@ end.
g(X) -> f(X+1).
\end{lstlisting}
\end{minipage}% This must go next to `\end{minipage}`
\quad{$\xrightarrow[\textit{(change expression)}]{\makebox[150pt]{\textit{wrap expression}}}$}

\noindent
{\Large$\rightarrow$}
\begin{minipage}{.59\textwidth}
\begin{lstlisting}[frame=single,xleftmargin=3pt,framexleftmargin=0pt]
f(X) -> @begin X*fun() -> 2 end() end@.
g(X) -> f(X+1).
\end{lstlisting}
\end{minipage}% This must go next to `\end{minipage}`
\quad{$\xrightarrow[\textit{(introduce function)}]{\makebox[150pt]{\textit{extract expression to function (z)}}}$}

\noindent
{\Large$\rightarrow$}
\begin{minipage}{.59\textwidth}
\begin{lstlisting}[frame=single,xleftmargin=3pt,framexleftmargin=0pt]
z(X) -> begin X*@fun() -> 2 end@() end.
f(X) -> z(X).
g(X) -> f(X+1).
\end{lstlisting}
\end{minipage}% This must go next to `\end{minipage}`
\quad{$\xrightarrow[\textit{(introduce variable)}]{\makebox[150pt]{\textit{extract expression to variable (Y)}}}$}

\noindent
{\Large$\rightarrow$}
\begin{minipage}{.59\textwidth}
\begin{lstlisting}[frame=single,xleftmargin=3pt,framexleftmargin=0pt]
z(X) -> begin @Y@=fun() -> 2 end, X*Y() end.
f(X) -> z(X).
g(X) -> f(X+1).
\end{lstlisting}
\end{minipage}% This must go next to `\end{minipage}`
\quad{$\xrightarrow[\textit{(change variable scope)}]{\makebox[150pt]{\textit{outer variable}}}$}

\noindent
{\Large$\rightarrow$}
\begin{minipage}{.59\textwidth}
\begin{lstlisting}[frame=single,xleftmargin=3pt,framexleftmargin=0pt]
@z@(X) -> Y=fun() -> 2 end, begin X*Y() end.
f(X) -> z(X).
g(X) -> f(X+1).
\end{lstlisting}
\end{minipage}% This must go next to `\end{minipage}`
\quad{$\xrightarrow[\textit{(change function)}]{\makebox[150pt]{\textit{variable to function parameter (Y)}}}$}

\noindent
{\Large$\rightarrow$}
\begin{minipage}{.59\textwidth}
\begin{lstlisting}[frame=single,xleftmargin=3pt,framexleftmargin=0pt]
@z@(X, Y) -> begin X*Y() end.
f(X) -> z(X, fun() -> 2 end).
g(X) -> f(X+1).
\end{lstlisting}
\end{minipage}% This must go next to `\end{minipage}`
\quad{$\xrightarrow[\textit{(change function signature)}]{\makebox[150pt]{\textit{rename function (z to f)}}}$}

\noindent
{\Large$\rightarrow$}
\begin{minipage}{.595\textwidth}
\begin{lstlisting}[frame=single,xleftmargin=3pt,framexleftmargin=0pt]
f(X, Y) -> begin X*Y() end.
f(X) -> f(X, fun() -> 2 end).
g(X) -> f(X+1).
\end{lstlisting}
\end{minipage}% This must go next to `\end{minipage}`

\noindent
After performing 6 small refactoring steps, we arrive at the same result we had in our example presented in the previous section, which is the core idea behind micro-refactoring. In the following section, we are going to define each of these refactoring transformations in our specification formalism, and we also define a composite refactoring function that controls the application of these constituent steps.

\subsection{Formal definition}
\label{sec:def}

\lstset{
  language=Erlang,
  frame=l,
  rulecolor=\color{lightgray},
  basicstyle=\small\ttfamily,
  keywordstyle=\bf,
  numbers=none,
  tabsize=4,
  breaklines=true,
  showspaces=false,
  showtabs=false,
  captionpos=b,
  morekeywords={REFACTORING,SELECTOR,DO,
                AND,OR,NOT,
                FORALL,FOR,WHEN,ITERATE,
                DEFINITION,REFERENCE,
                LOCAL, INTRODUCE, VARIABLE, FUNCTION, SIGNATURE,
                IN, OUTER, SCOPE, MODULE,
                THEN, ON, THIS},
}

In this section, we give a formal specification for the steps used in the decomposition. Our definition includes two composite function definitions and six prime refactoring functions derived from multiple schemes. We omit the formal definition of the selector ``function\_part'', which queries the lambda function from the result of wrapping, and we also use some semantic functions that are not defined formally in this presentation.

For the syntax and semantics of the refactoring formalism, we refer to the comprehensive introduction presented in~\cite{VPT16}. Roughly summarising, composite functions define a refactoring sequence, whilst prime functions instantiate schemes with rewrite rules applied to the definition and the references of some data or name. In any refactoring function, ``THIS'' is a local variable that holds a reference of the object of interest, that is, the program element the refactoring function is applied to (the refactoring target).

\vspace{-0.5em}
\subsubsection{Composite refactoring functions}

The main refactoring function is called \texttt{generalise\_function} and it takes one argument determining the name of the new variable added to the function signature. It is merely a sequential composition of the rest of the refactoring functions, though it refers to two semantic functions as well: \texttt{function} associates the containing function with any syntactic element, while \texttt{name} simply returns the name of the function.

\begin{lstlisting}
REFACTORING generalise_function(ParamName)
DO
    THIS.wrap()
    THIS = THIS.function_part()
    Old = function(THIS)
    Name = name(Old)
    Params = function_params(Old)
    New = Old.body().extract_to_function(tmp, Params)
    Var = THIS.extract_to_variable(ParamName)
    Var.to_function_parameter()
    New.rename_function(Name)
\end{lstlisting}

\noindent
The result of one transformation can be the target or argument to other functions, like in pipelines. When a component step fails, the entire composition fails, and all intermediate results are rolled back. Although incomplete composite functions are correct as all composed steps are refactorings alone, intermediate changes may be undesired.

We introduce another composite refactoring function, \texttt{to\_function\_parameter}, which is targeting the variable matching created by a preceding step, and it lifts the new variable into a function parameter. This involves two different steps: it iterates lifting between scopes until the variable reaches the scope of the function (at this point the iteration construct will terminate successfully), and then it lifts the function-level variable to the parameter list.

\begin{lstlisting}
REFACTORING to_function_parameter()
DO
    ITERATE THIS.outer_variable()
    function(THIS).var_to_param(THIS)
\end{lstlisting}

\noindent
Note that by resembling method invocation in object-oriented languages, we allow refactoring functions be applied to language elements specified either with a variable or by using selector functions.

\subsubsection{Prime refactoring functions}

Prime refactorings implement micro-refactorings. They are expressed with refactoring schemes that control the application of the rewrite rules used for scheme instantiation. We define a local refactoring for making an arbitrary expression ``movable'', as a result of which the expression is wrapped into a lambda function. Though being a simple change, it has complex conditions: it requires that the expression does not bind any variables that are used outside the expression (predicate \texttt{non\_bind}), while another condition binds a metavariable to hold the variable names that are free in the expression (referred to by the expression, but bound in the context).

\begin{lstlisting}
LOCAL REFACTORING wrap()
    E
    -------------------------------
    (fun(Vars..) -> E end) (Vars..)
WHEN
    Vars.. = free_vars(E) AND non_bind(E)
\end{lstlisting}

\noindent
We define two instances of the variable introduction scheme for introducing and lifting the new parameter of the generalised function. Both define the syntactic construct creating the binding, determine the place (the scope) of the binding, and they also specify the rewrite rule that will transform the target expression to use the newly introduced binding. Since new variable bindings can be placed either in the current scope or in an outer scope, this has to be decided in the instantiation of the scheme, while the conditions regarding name clash should be handled inherently. With this, we can express both introduction and lifting with the same scheme, and in the second one, the variable name is coming from the already present binding rather than from the refactoring argument. Syntactic noise (e.g. dead scope) introduced by intermediate steps can be removed by dedicated clean-up refactorings at the end of the process, however, we do not include clean-up transformations in this presentation.

\noindent
\begin{minipage}{.45\textwidth}
\begin{lstlisting}
INTRODUCE VARIABLE
  extract_to_variable(Name)
DEFINITION IN SCOPE
    Name = E
REFERENCE
    E
    ----
    Name
\end{lstlisting}
\end{minipage}% This must go next to `\end{minipage}`
\quad
\begin{minipage}{.45\textwidth}
\begin{lstlisting}
INTRODUCE VARIABLE
  outer_variable()
DEFINITION IN OUTER SCOPE
    Name = E
REFERENCE
    Name = E
    --------
    Name
\end{lstlisting}
\end{minipage}% This must go next to `\end{minipage}`

\noindent
We use the function introduction scheme for creating the fall-back function. Unlike in variable introduction, function definition placement is not an issue (the module name space is flat in Erlang), it does not matter where in a module a function is placed. The scheme implementation will append the new definition to the file.

\begin{lstlisting}
INTRODUCE FUNCTION extract_to_function(Name, Params..)
DEFINITION
    Name(Params..) -> E .
REFERENCE
    E
    --------------
    Name(Params..)
WHEN is_subset(free_vars(E), vars(Params..))
\end{lstlisting}

\noindent
Perhaps the most interesting components in the compound refactoring are the ones transforming the function and its signature. The function refactoring scheme transforms the function as well as its references by applying the supplied rewrite rules on the definition and on all kinds of references, including calls, name references and directives.

\begin{lstlisting}
FUNCTION REFACTORING var_to_param(X)
DEFINITION
    (Args..) -> X = E, Body..
    -------------------------
    (Args.., X) -> Body..
REFERENCE
    (Args2..)
    ------------
    (Args2.., E)
WHEN pure(E) AND closed(E)
\end{lstlisting}

\noindent
A special case of function refactoring is function signature refactoring, which only transforms the head of the function definition and its refereces. As we demonstrated in our previous paper~\cite{VPT16}, this scheme can be used as well for renaming a function and to restructure or reorder its arguments.

\begin{lstlisting}
FUNCTION SIGNATURE REFACTORING rename_function(NewName)
    Name(Args..)
    ---------------
    NewName(Args..)
\end{lstlisting}

\section{Verification of the case study}
\label{sec:verification}

In this section we demonstrate the formal verification of some components of our case study defined in Section~\ref{sec:def}. In order to do so, we incorporate our definition of formal semantics and formula of semantic equivalence designed for our example object language, Erlang.

\subsection{Equivalence of Erlang expression patterns}

As mentioned already, correctness of a refactoring definition is reduced to an equivalence problem of expression patterns, so at first, we need to find a sensible and appropriate definition of semantic equivalence between Erlang expression patterns. Since semantic equivalence can be defined in different granularities (determined by the details we take into account about the program elements and their execution environment), the way we formalise equivalence has a great influence on what patterns will be regarded as being equal. For instance, equivalence can be defined to only compare the values of expressions, but it might also expect the expressions to have the same series of side effects or to show the same behaviour on exceptional inputs. Although the configuration in our object language semantics is rather fine-grained (i.e. it captures module structure with module attributes and function definitions, variable environment and side effects), in our proofs we abstract on this and make equivalence requirements only on the code part and its variable environment. Also, for some proofs, this has to be relaxed to include the side effects:
\[
\cell{cfg}{\cell{code}{}\; \cell{env}{} \; \cell{side\_eff}{}}
\]

Equivalence has to imply the following property: the expressions are interchangeable in any program context so that the meaning of the entire program is preserved. The equivalence specification of Erlang expression patterns should ensure that any concrete expression pairs instantiated from the patterns are evaluated to the same value, their evaluation has the same impact on the variable environment, and the same side effects occur during their evaluation.

This equivalence relation can be specified with matching logic formulas. Matching logic allows us to specify patterns over program configurations, and additional constraints can be added to the configuration expressed with regular first-order logic expressions. For equivalence checking, the configuration of the equivalence problem is composed of two configurations of the object programming language. We specify equivalence of Erlang program patterns with two formulas: one regarding successful evaluation and one for failure.

The first formula says that the two patterns can be rewritten to the same form (are 'joinable'), side effects occurred in the same number and order, and variable mappings are identical, too. The second formula matches pairs of program states where both executions terminate with an exception (the types of exceptions do not have to be the same, though). Unlike Erlang, the sub-language we formalized does not provide exception handling, in our model exceptions will simply terminate the execution and ignore the rest of the program.

\smsnn{
  \term{Eq} \equiv \left\{
  \bigcell{eq}{
  \begin{aligned}  
   \cell{cfg}{
      &\cell{code}{\var{Code}} 
      &&\cell{env}{\varepsilon}
      &\cell{side\_eff}{\var{SE}}} \\
   \cell{cfg}{
      &\cell{code}{\var{Code}} 
      &&\cell{env}{\varepsilon}
      &\cell{side\_eff}{\var{SE}}} \\
   \end{aligned}} ,
  \bigcell{eq}{
  \begin{aligned}  
   \cell{cfg}{
      &\cell{code}{} 
      &&\cell{env}{\varepsilon_1}
      &\cell{side\_eff}{\var{SE}, \term{\#exception}}} \\
   \cell{cfg}{
      &\cell{code}{} 
      &&\cell{env}{\varepsilon_2}
      &\cell{side\_eff}{\var{SE}, \term{\#exception}}} \\
   \end{aligned}}
   \right\}
}

\subsection{Proof system}

The general-purpose reachability proof system, referred to in our previous publication~\cite{VPT16}, can generate proofs automatically, but it can only prove partial equivalence and not full equivalence. Moreover, this generic proof system generates unnecessarily many branches, because it cannot take into account the specialities of equivalence proofs of deterministic programs. Since then, a new matching logic based language independent 5-rule proof system was published for checking full program equivalence of deterministic programs~\cite{full-eq}. The inference rules are directly designed for equivalence proofs, making the reasoning higher-level, effective and readable. Moreover, it can also prove equivalence of non-terminating programs. Unfortunately, there is no tool support for this new Equivalence Proof System yet, but most formulas can still be automatically proved with the implementation of the general-purpose proof system.

In this paper, we apply the inference rules of the new, equivalence-focussed proof system, as it provides a more appropriate toolkit for phrasing our correctness checks. We are going to prove formulas of form $\vdash \varphi \Downarrow^{\infty} \term{Eq}$, which express the derivability of an equivalence formula $\varphi$ in the proof system with respect to a fixed formula set $\term{Eq}$ characterizing the pairs of programs that are known to be equivalent. The formula is validated by applying inference rules of the proof system in the traditional bottom-up manner until every leaf of the proof tree is an instance of the \textsc{Axiom} rule. For reasoning about our use case, we use the following three rules out of the five:

\smsnn{
\begin{aligned}
\textsc{Axiom} \;
\frac
{\varphi \in  \term{Eq}}
{\vdash \varphi \Downarrow^{\infty} \term{Eq}}
\;\;\;\;\;\;\;\;
\textsc{Conseq} \;
\frac
{\vDash \varphi \rightarrow \exists \tilde{x}.\varphi' \;\;\;\;
 \vdash \varphi' \Downarrow^{\infty}  \term{Eq}}
{\vdash \varphi \Downarrow^{\infty}  \term{Eq}}
\\
\\
\textsc{Step} \;
\frac 
{\vDash \varphi_1 \Rightarrow^*_1 \varphi'_1 \;\;\;\;
 \vDash \varphi_2 \Rightarrow^*_2 \varphi'_2 \;\;\;\;
 \vdash \cell{}{\varphi'_1, \varphi'_2} \Downarrow^{\infty}  \term{Eq}}
{\vdash \cell{}{\varphi_1, \varphi_2} \Downarrow^{\infty}  \term{Eq}}
\end{aligned}
}

\textsc{Axiom} states that every formula of the set $\term{Eq}$ is derivable. \textsc{Conseq} allows to perform domain reasoning and prove a more general formula instead of the specific one. Last but not least, \textsc{Step} allows to take arbitrary number of finite steps in each of the two program configurations. This rule weaves operational semantic rules into the reachability reasoning. For further details, see~\cite{full-eq}.

\subsection{Proof sketch for the local refactoring}

The correctness property for a local refactoring can be expressed by the equivalence problem of the two expression patterns given in the rewrite rule, extended with the conditions. This formula can be mechanically constructed, by inspecting the conditional rewrite rule defining the refactoring:

\begin{lstlisting}
    <pattern1>
    -----------  WHEN <condition>
    <pattern2>
\end{lstlisting}

\noindent
The corresponding formula captures the equivalence of the patterns according to our previous definition of expression pattern equivalence. The condition is simply attached to the configuration pattern:
\smsnn{
\varphi \equiv
\cell{eq}{\cell{cfg}{\cell{code}{\term{<pattern1>}} \cell{env}{\varepsilon}}\;
           \cell{cfg}{\cell{code}{\term{<pattern2>}} \cell{env}{\varepsilon}}}
           \wedge \term{<condition>}
}

\noindent
Now let us demonstrate how the \textit{wrap} local refactoring is verified in the proof system. First, let $\Psi$ be the condition of the refactoring rule: $\Psi \equiv \var{Vars} = \term{free\_vars}(\var{E}) \wedge \term{non\_bind}(\var{E})$. Note that italic upper-case names (like $\var{E}$) denote mathematical variables rather than program variables. So-called list metavariables (postfixed by ``\verb|..|'', see~\cite{VPT16}) become regular variables of the formula (e.g. $\var{Vars}$), because in matching logic we can capture them with types of sequences of elements (e.g. $\var{VarList}$). In the formula, a variable followed by a colon and a type name is a type-restricted variable; by default, all variables have the most general type allowed by the syntactic context. For explanation on the ``followed by'' ($\kra$) and the ``hole'' ($\khole$) symbols we refer to the comprehensive overview of the $\mathbb{K}$ framework~\cite{koverview}.

The proof~\eqref{eq:proof1} is read bottom up starting with line 3 (the proof goal). We perform \textsc{Step} by using semantic rules and semantic lemmas on the second configuration of the goal, which leads to state 2. Then, by generalising the formula, we get the axiom: one of the equivalence specification formulas. Note that the cell of side effects is hidden as it remains empty during the proof.

\sms[eq:proof1]{
\begin{aligned}
& 1.
  && \vdash \bigcell{eq}{
  \begin{aligned}  
   \cellpre{cfg}{
      &\cell{code}{\var{Code}} 
      &&\cell{env}{\varepsilon}}\\
   \cellpre{cfg}{
      &\cell{code}{\var{Code}} 
      &&\cell{env}{\varepsilon}}\\
   \end{aligned}}
  &\Downarrow^{\infty} && \term{Eq} && \textsc{Axiom} \\
& 2.
  && \vdash \bigcell{eq}{
  \begin{aligned}  
   \cellpre{cfg}{
      &\cell{code}{\var{E}} 
      &&\cell{env}{\varepsilon}} \\
   \cellpre{cfg}{
      &\cell{code}{\var{E}} 
      &&\cell{env}{\varepsilon}} \\
   \end{aligned}}\; \wedge\; \Psi
  &\Downarrow^{\infty} && \term{Eq} && \textsc{Conseq(1)} \\
& 3. 
  && \vdash \bigcell{eq}{
  \begin{aligned}  
   \cellpre{cfg}{
      &\cell{code}{\var{E}} 
      &&\cell{env}{\varepsilon}} \\
   \cellpre{cfg}{
      &\cell{code}{\term{fun}(\var{Vars}) \term{->} \var{E} \term{end} (\var{Vars})} 
      &&\cell{env}{\varepsilon}} \\
   \end{aligned}}\; \wedge\; \Psi
 &\Downarrow^{\infty} && \term{Eq} && \textsc{Step(2)}
\end{aligned}
}\vspace{-0.5em}

To validate the application of the \textsc{Step} rule, we need to prove the following reachability:
\smsnn{
\cellpre{cfg}{
  \cell{code}{\term{fun}(\var{Vars}) \term{->} \var{E} \term{end} (\var{Vars})} 
   \cell{env}{\varepsilon}}  \wedge \Psi
\; \Rightarrow^* \;
\cellpre{cfg}{
    \cell{code}{\var{E}} 
    \cell{env}{\varepsilon}} \wedge \Psi
}\vspace{-1em}

Starting with the left-hand-side configuration of the reachability, we rewrite the current state by applying either a proper semantic rule or a lemma until we get the right-hand-side configuration:

\smsnn{
\begin{aligned}
& \term{cfg}_{0} &&:&&
\cellpre{cfg}{
  \cell{code}{\term{fun}(\var{Vars}) \term{->} \var{E} \term{end} (\var{Vars})} 
   \cell{env}{\varepsilon}}  \wedge \Psi \\
& \text{rule}_1 &&:&& 
\cellpre{cfg}{\cellpre{code}{\varType{F}{Fun}(\var{Exp})}} \wedge \neg \term{isValue}(\var{Exp})
\Rightarrow 
\cellpre{cfg}{\cellpre{code}{\var{Exp} \kra \var{F}(\khole)}} \\
%\\
& \term{cfg}_{1} &&: &&
   \cellpre{cfg}{
      \cell{code}{\var{Vars} \kra \term{fun}(\var{Vars}) \term{->} \var{E} \term{end} (\khole)}
      \; \cell{env}{\varepsilon}} \wedge \Psi \\
& \text{rule}_2 &&:&& 
\cellpre{cfg}{\cellpre{code}{\varType{Vs}{VarList}}}
\Rightarrow 
\cellpre{cfg}{\cellpre{code}{\term{lookup}(\var{Vs})}} \\
%\\
& \term{cfg}_{2} &&: &&
    \cellpre{cfg}{
      \cell{code}{\term{lookup}(\var{Vars},\varepsilon) \kra \term{fun}(\var{Vars}) \term{->} \var{E} \term{end} (\khole)} 
      \; \cell{env}{\varepsilon}} \wedge \Psi \\
& \text{rule}_3 &&:&& 
\cellpre{cfg}{\cellpre{code}{\varType{V}{Value} \kra \varType{F}{Fun}(\khole)}}
\Rightarrow 
\cellpre{cfg}{\cellpre{code}{\varType{F}{Fun}(\var{V})}} \\
%\\
& \term{cfg}_{3} &&: &&
     \cellpre{cfg}{
      \cell{code}{\term{fun}(\var{Vars}) \term{->} \var{E} \term{end} (\term{lookup}(\var{Vars},\varepsilon))} 
      \; \cell{env}{\varepsilon}} \wedge \Psi\\
& \text{rule}_4 &&:&& 
\cellpre{cfg}{\cellpre{code}{\term{fun} (\varType{Ps}{PatternList}) \term{->} \var{E} \term{end}(\varType{V}{Value})} \cell{env}{\varepsilon}}
\\  
&&&&& \Rightarrow \cellpre{cfg}{\cellpre{code}{\term{matches}(\var{V}, \var{Ps} \term{->} \var{E}) \kra \term{restoreEnv}(\varepsilon)} \cell{env}{\term{remove}(\term{vars}(\var{Ps}),\varepsilon)}} \\
%\\
& \term{cfg}_{4} && : &&
     \cellpre{cfg}{
      \cell{code}{\term{matches}(\term{lookup}(\var{Vars},\varepsilon), \var{Vars} \term{->} \var{E})\;\kra\; \term{restoreEnv}(\varepsilon)} \\
      &&&&&\;\;\cell{env}{\term{remove}(\var{Vars},\varepsilon)}} \wedge \Psi\\
& \text{rule}_5 &&:&& 
\cellpre{cfg}{\cellpre{code}{\term{matches}(\var{V}, \var{Ps} \term{->} \var{E})} \cell{env}{\varepsilon}} \wedge \term{isMatching}(\var{V}, \var{Ps}, \varepsilon) 
\\ &&&&& \Rightarrow \cellpre{cfg}{\cellpre{code}{\var{E}} \cell{env}{\varepsilon \term{++} \term{getMatching}(\var{V}, \var{Ps}, \varepsilon)}} \\      
%\\    
& \term{cfg}_{5} && : &&
     \cellpre{cfg}{
      \cell{code}{\var{E}\;\kra\; \term{restoreEnv}(\varepsilon)} \\
      &&&&&\;\;\cell{env}{\term{remove}(\var{Vars},\varepsilon) \term{++} \term{getMatching}(\term{lookup}(\var{Vars},\varepsilon), \var{Vars}, \term{remove}(\var{Vars},\varepsilon))}} \wedge \Psi\\
& \text{lemma}_1 &&:&& 
\term{remove}(\var{Vars},\varepsilon) \term{++} \term{getMatching}(\term{lookup}(\var{Vars},\varepsilon), \var{Vars}, \term{remove}(\var{Vars},\varepsilon))
\Rightarrow
\varepsilon \\
%\\
& \term{cfg}_{6} && : &&
     \cellpre{cfg}{
      \cell{code}{\var{E}\;\kra\; \term{restoreEnv}(\varepsilon)} \;
      \cell{env}{\varepsilon}} \wedge \Psi\\
& \text{lemma}_2 &&:&& 
\cellpre{cfg}{
  \cell{code}{\var{E}\;\kra\; \term{restoreEnv}(\varepsilon)} \;
  \cell{env}{\varepsilon}} \wedge \term{non\_bind}(\var{E})
\Rightarrow
\cellpre{cfg}{
  \cell{code}{\var{E}} \;
  \cell{env}{\varepsilon}} \\  
%\\
& \term{cfg}_{7} && : &&
     \cellpre{cfg}{
      \cell{code}{\var{E}} \;
      \cell{env}{\varepsilon}} \wedge \Psi  
\end{aligned}
}

Let us give a brief explanation for the semantic proof.

\begin{itemize}
\item Initially, we have a call for an anonymous function. The actual parameters of the call have to be evaluated to a value before the call itself is evaluated, so with $\term{rule}_1$ we force the evaluation of the parameters by moving them to the top of the cell (``heating'' rule).
\item The $\var{Vars} = \term{free\_vars}(\var{E})$ statement of the condition $\Psi$ implies that the type of the $\var{Vars}$ is list of variables. Variables are evaluated by looking up their associated values from the variable environment, by using the function $\term{lookup}$ ($\term{rule}_2$). Note that $\term{lookup}$ is not evaluated (the variables as well as the environment are symbolic), but we know that its return value has the type $\var{Value}$, so we can put it back to the argument of the call applying $\term{rule}_3$ (``cooling'' rule).

\item Then, $\term{rule}_4$ rewrites the function call to the $\term{matches}$ intermediate structure used for program constructs where values have to be matched against patterns. The formal parameters of the anonymous function may shadow variables defined in the outer scope; therefore, all variables occurring in the patterns of the parameter list are removed from the variable environment. Besides, as the anonymous function opens a new scope, variable bindings inside the function should not have any effects on the containing scope. To reflect this, an extra operation is inserted into the \verb|code| cell: $\term{restoreEnv}$ is responsible for restoring the original variable environment after the call has been fully evaluated.

\item We can apply $\term{rule}_5$ in the case when the value matches the pattern of the clause of the $\term{matches}$ structure. In our case, the $\term{isMatching}$ function of the rule constraint is evaluated to \verb|true|, as $\var{Vars}$ contains only variables that do not occur in the current variable environment (because of the $\term{remove}$ operation), so they will match any value. Moreover, the length of the value list and the variable list are the same, because the $\term{lookup}$ function does not change the length of the list given as parameter.

\item Using $\term{lemma}_1$ we simplify the variable environment. The lemma can be derived from the definitions of contained functions. Informally, it states that if we remove variable assignments from the environment, and then we put back the same assignments, we obtain the original environment. Finally, $\term{lemma}_2$ expresses the fact that if we need to restore a variable environment after evaluating an arbitrary expression that does not introduce variable bindings, we can omit the $\term{resotreEnv}$ operation containing the current environment in its parameter.
\end{itemize}

\subsection{Proof sketch for a scheme instance}

As clarified already, verification of a refactoring expressed as a scheme instance is two-phased: 1) the scheme has to be verified w.r.t. its contract on its instantiation, and 2) the instantiation has to be checked for conformance with the contract. In this section, we demonstrate the verifiability of the ``extract to variable'' refactoring function.

\subsubsection*{The variable introduction scheme}

This scheme is characterized by the binding it creates (the name it introduces with the expression the name is bound to), and the rewrite rule that transforms the originally selected expression. When defining an instance of this scheme, the refactoring programmer, according to a pre-defined format, precisely specifies these elements. In this specific case, the introduction place is fixed to be the scope of the selected expression.

\newpage

\noindent
The generic format of the scheme is the following:

\begin{lstlisting}
DEFINITION IN SCOPE
    <name> = <pattern1>
REFERENCE
    <pattern2>
    ----------
    <pattern3>
\end{lstlisting}

\noindent
The contract for this scheme requires that
\[
\texttt{begin <name> = <pattern1>, <pattern3> end} \quad \equiv \quad \texttt{<pattern2>}
\]
\noindent under the side-conditions specified in the unfolding of the scheme.

It is apparent that the scheme can only be used for introducing variable bindings, but it also allows for modification of the reference site. According to the operational definition of the scheme, it is transformed into an extensive strategy relying on the semantic relation of scopes:

\begin{lstlisting}
ON scope(THIS)
  E..
  ------------------------
  <name> = <pattern1>, E..
WHEN fresh(<name>) AND pure(<pattern1>) AND closed(<pattern1>)
THEN ON THIS
  <pattern2>
  ----------
  <pattern3>
\end{lstlisting}

This extensive scheme will transform the selected expression (``ON THIS'') as well as its scope in order to carry out a consistent change. The expression is certainly a sub-expression of its scope, but the structural correspondence between them is defined by the semantic function ``scope'' --- the transformation as a whole can be seen as a rewriting that involves a special semantic pattern. In order to prove the correctness of this extensive rewriting, we need to check that in any environment, with any particular nesting of the expression, the transformation is going to result in semantically equivalent code.

In order to cover every possible case, we rephrase the semantic relation as an inductively defined set by building on syntactic relations. For the sake of simplicity, let us now define \emph{scope(X)} as the containing block of the expression, formally, \textit{block(top\_expression(X))}, where the relation \textit{top\_expression} is inductively defined with the direct sub-expression relation as its base case. Then we do the proof by induction on the structure of the relation \emph{scope}, which in our case is a structural induction on the syntactic relationship between the expression and its containing block. Roughly speaking, we need to simulate all the possible embeddings of the expression in the block it is located inside, and compare these with their variants transformed by the instantiation rules.

The base case for this structural induction proof is characterised by the ``empty environment'', meaning the expression is a singleton top-level element of the block. Fortunately, this case is completely justified by the contract of the scheme. Then, by considering every single case where the expression is a sub-expression of another (characterising the inductive step of top\_expression), we prove that regardless of the structure in the block, the equivalence relation holds. This step can be easily carried out based on the operational semantic rules that cover the entire language.

\subsubsection*{The scheme instance}

In order to verify the scheme instance, we compose the equivalence formula by inspecting the instantiation rules. As for the ``extract to variable'' function, the following formula will express the validity of the scheme instance:

\smsnn{
\begin{aligned}
  && \vdash  \bigcell{eq}{
  \begin{aligned}  
   \cellpre{cfg}{
      &\cell{code}{\var{E}}
      &&\cell{env}{\varepsilon}} \\
   \cellpre{cfg}{
      &\cell{code}{\term{begin} \var{Name} = \var{E}, \var{Name} \term{end}} 
      &&\cell{env}{\varepsilon}} \\
   \end{aligned}}
 \wedge\; \text{fresh}(\var{Name})
 &\Downarrow^{\infty} && \term{Eq}
\end{aligned}
}

\noindent
The formula can be proved with the same inference rules as the example proof we showed for the \textit{wrap} refactoring.

\section{Conclusion}
\label{sec:conc}

We have presented a verifiable yet executable definition of a fairly complex refactoring transformation, aiming at demonstrating the applicability of our approach to refactoring formalisation. We have decomposed ``generalise function'', a well-known refactoring, in an unforeseen way into a couple of small semantics-preserving steps, each of which are semi-automatically verifiable. Our case study has justified that by decomposing composite transformations into instances of several simple refactoring scheme instances, refactorings of various complexities may be expressed and defined in a verifiable manner.

The core ideas of decomposition and refactoring schemes have been demonstrated for the Erlang programming language, but at the same time, major results of our work have been shown to be mostly language-agnostic and applicable to abstractions of other programming paradigms. In the future, we intend to compose a large number of refactoring definitions for a wide range of use cases and document their formal proof in full detail. Besides, we would like to define the refactoring language formally, both the operational behaviour and the method for turning refactoring definitions into automatically verifiable formulas.

\section{Acknowledgement}
\label{sec:ack}

The authors would like to express their sincere gratitude to Prof. Simon Thompson for his help and support in guiding the first steps in this research. We thank the reviewers for their constructive and helpful comments.

This research was supported through the New National Excellence Program of the Ministry of Human Capacities of Hungary.

% ------------------------------------------------------------------------------
\label{sect:bib}
\bibliographystyle{eptcs}
\bibliography{vpt17}
% ------------------------------------------------------------------------------

\end{document}